\documentclass[useAMS,usenatbib]{mn2e}
\usepackage{times}
\usepackage{graphicx}
\usepackage{subfigure}
\usepackage{psfrag}
\usepackage{amsmath}
\usepackage{amssymb}

\def\reff@jnl#1{{\rm#1\/}}
\def\aj{\reff@jnl{AJ}}                 % Astronomical Journal
\def\araa{\reff@jnl{ARA\&A}}           % Annual Review of Astron and Astrophys
\def\apj{\reff@jnl{ApJ}}               % Astrophysical Journal
\def\apjl{\reff@jnl{ApJ}}              % Astrophysical Journal, Letters
\def\apjs{\reff@jnl{ApJS}}             % Astrophysical Journal, Supplement
\def\ao{\reff@jnl{Appl.Optics}}        % Applied Optics
\def\apss{\reff@jnl{Ap\&SS}}           % Astrophysics and Space Science
\def\aap{\reff@jnl{A\&A}}              % Astronomy and Astrophysics
\def\aapr{\reff@jnl{A\&A~Rev.}}        % Astronomy and Astrophysics Reviews
\def\aaps{\reff@jnl{A\&AS}}            % Astronomy and Astrophysics, Supplement
\def\azh{\reff@jnl{AZh}}               % Astronomicheskii Zhurnal
\def\baas{\reff@jnl{BAAS}}             % Bulletin of the AAS
\def\jrasc{\reff@jnl{JRASC}}           % Journal of the RAS of Canada
\def\memras{\reff@jnl{MmRAS}}          % Memoirs of the RAS
\def\mnras{\reff@jnl{MNRAS}}           % Monthly Notices of the RAS
\def\pra{\reff@jnl{Phys.Rev.A}}        % Physical Review A: General Physics
\def\prb{\reff@jnl{Phys.Rev.B}}        % Physical Review B: Solid State
\def\prc{\reff@jnl{Phys.Rev.C}}        % Physical Review C
\def\prd{\reff@jnl{Phys.Rev.D}}        % Physical Review D
\def\prl{\reff@jnl{Phys.Rev.Lett}}     % Physical Review Letters
\def\pasp{\reff@jnl{PASP}}             % Publications of the ASP
\def\pasj{\reff@jnl{PASJ}}             % Publications of the ASJ
\def\qjras{\reff@jnl{QJRAS}}           % Quarterly Journal of the RAS
\def\skytel{\reff@jnl{S\&T}}           % Sky and Telescope
\def\solphys{\reff@jnl{Solar~Phys.}}   % Solar Physics
\def\sovast{\reff@jnl{Soviet~Ast.}}    % Soviet Astronomy
\def\ssr{\reff@jnl{Space~Sci.Rev.}}    % Space Science Reviews
\def\zap{\reff@jnl{ZAp}}               % Zeitschrift fuer Astrophysik
\def\nat{\reff@jnl{Nature}}            % Nature
\title[Analysis of RV data with red noise] {Bayesian analysis of radial velocity data of GJ667C with correlated noise: evidence for only 2 planets} 
\author[F.~Feroz and
  M.P.~Hobson] {F.~Feroz\thanks{E-mail: f.feroz@mrao.cam.ac.uk} and M.~P.~Hobson\\Astrophysics Group, Cavendish
  Laboratory, JJ Thomson Avenue, Cambridge CB3 0HE, UK\\}

\date{Accepted ---. Received ---; in original form \today}
\pagerange{\pageref{firstpage}--\pageref{lastpage}}
\pubyear{2013}

\begin{document}
\label{firstpage}
\maketitle

\begin{abstract}
GJ667C is the least massive component of a triple star system which
lies at a distance of about $6.8$ pc (22.1 light-years) from
Earth. GJ667C has received much attention recently due to the claims
that it hosts up to seven planets including three super-Earths inside
the habitable zone. We present a Bayesian technique for the analysis
of radial velocity (RV) data-sets in the presence of correlated noise
component (``red noise''), with unknown parameters. We also introduce
hyper-parameters in our model in order to deal statistically with
under or over-estimated error bars on measured RVs as well as
inconsistencies between different data-sets. By applying this method
to the RV data-set of GJ667C, we show that this data-set contains a
significant correlated (red) noise component with correlation
timescale for HARPS data of order $9$ days. Our analysis shows that
the data only provides strong evidence for the presence of two
planets: GJ667Cb and c with periods $7.19$d and $28.13$d respectively,
with some hints towards the presence of a third signal with period
$91$d. The planetary nature of this third signal is not clear and 
additional RV observations are required for its confirmation. Previous
claims of the detection of additional planets in this system are due
the erroneous assumption of white noise. Using the standard white
noise assumption, our method leads to the detection of up to five
signals in this system. We also find that with the red noise model, 
the measurement uncertainties from HARPS for this system are 
under-estimated at the level of $\sim 50$ per cent.

\end{abstract}

\begin{keywords}
stars: planetary systems -- stars: individual: GJ667C -- techniques: radial velocities -- methods: data analysis -- methods: statistical
\end{keywords}

%%%%%%%%%%%%%%%%%%%%%%%%%%%%%%%%%%%%%%%%%%%%%%%%%%%%%%%%%
\section{Introduction}\label{sec:intro}
%%%%%%%%%%%%%%%%%%%%%%%%%%%%%%%%%%%%%%%%%%%%%%%%%%%%%%%%%

Extrasolar planetary research has made great advances in the last
decade as a result of the data gathered by several ground and space
based telescopes and thus far more than 900 extrasolar planets have been
discovered. More and more planets with large orbital periods and small
velocity amplitudes are now being detected due to remarkable
improvements in the accuracy of RV measurements. With the flood of new
data, more powerful statistical techniques are being developed and
applied to extract as much information as possible. Traditionally, the
orbital parameters of the planets and their uncertainties have been
obtained by a two stage process. First the period of the planets is
determined by searching for periodicity in the RV data using the
Lomb--Scargle periodogram (\citealt{1976Ap&SS..39..447L,
  1982ApJ...263..835S}). Other orbital parameters are then determined
using minimisation algorithms, with the orbital period of the planets
fixed to the values determined by Lomb--Scargle periodogram.

Bayesian methods have several advantages over traditional methods, for
example when the data do not cover a complete orbital phase of the
planet. Bayesian inference also provides a rigorous way of performing
model selection which is required to decide the number of planets
favoured by the data. The main problem in applying such Bayesian model
selection techniques is the computational cost involved in calculating
the Bayesian evidence. Nonetheless, Bayesian model selection has the
potential to improve the interpretation of existing observational data
and possibly detect yet undiscovered planets. Recent advances in
Marko-Chain Monte Carlo (MCMC) techniques (see e.g. \citealt{MacKay})
have made it possible for Bayesian techniques to be applied to
extrasolar planetary searches (see e.g. \citealt{2005ApJ...631.1198G,
  2005AJ....129.1706F, 2007ASPC..371..189F,
  2009MNRAS.394.1936B}). \citet*{2011MNRAS.415.3462F} presented a new
Bayesian method for determining the number of extrasolar planets, as
well as for inferring their orbital parameters, without having to
calculate directly the Bayesian evidence for models containing a large
number of planets.

GJ667 is an M dwarf in a triple star system which lies at a distance of about $6.8$
pc (22.1 lightyears) from Earth. GJ667C is the least massive component
of this system with mass $0.33 \pm 0.03$ $M_{\sun}$
\citep{2013A&A...553A...8D}. Two other components of this system,
GJ667AB, are a closer couple of K dwarfs with semi-major axis of 1.82
AU, period of 42.15 years and mass of 1.27 $M_{\sun}$
\citep{1999A&A...341..121S}. GJ667C is at a projected distance of
$32.4''$ from GJ667AB, giving an expected semi-major axis of $\sim 300$ AU
\citep{2013A&A...553A...8D}. Using the data from the HARPS spectrograph
(with RVs obtained using cross-correlation function `CCF' technique),
\cite{2011arXiv1111.5019B} reported detection of a planet (GJ667Cb)
with orbital period of 7.2d and minimum mass of $5.9M_{\earth}$. They
also found evidence for the presence two further planets with orbital
periods 28d and 90d respectively. 7.2d and 28d planets were confirmed
by \cite{2012ApJ...751L..16A} and \cite{2013A&A...553A...8D}, both
using HARPS data although reduced using different techniques. GJ667Cc
with orbital period of 28d is particularly interesting as it lies well
within the habitable zone of the host star where it could support
liquid water. \cite{2012ApJ...751L..16A} further found evidence for
one additional planet with orbital period 75d, however they did not
consider it significant since it was affected by aliasing
interactions with another 91d signal and the likely rotation period
of the star at 105 days. \cite{2013A&A...553A...8D} found a signal
with orbital period 106d but attributed it to the stellar rotation due
to it being very close to the rotation period of the star. A Bayesian
analysis of the HARPS RVs for this system was performed by
\cite{2012arXiv1212.4058G}, who apart from confirming the presence of
first two planets GJ667Cb and c, also found evidence for four
additional signals with orbital periods 30.82d, 38.82d, 53.22d and
91.3d respectively. They discarded the 53.22d signal due to the high
likelihood of it being the second harmonic of the stellar rotational
period. Potential planets with 30.82d and 38.82d orbital periods lie
in the central region of their host star's habitable zone and
therefore are of much interest.

More recently, \cite{2013arXiv1306.6074A} performed a joint analysis
of RV observations of this system from HARPS, HIRES/Keck and
PFS/Magellan spectrographs (available from
\citealt{2012ApJS..200...15A}). Instead of using the CCF technique to
obtain RVs from observed spectra, they used the Template-Enhanced
Radial velocity Re-analysis Application `TERRA' technique which has
been claimed to produce significantly more accurate RVs compared to
RVs obtained using the CCF technique
(\citealt{2012ApJS..200...15A}). \cite{2013arXiv1306.6074A} found
evidence for the existence of six (even seven) planets GJ667Ca-f with
period 7.2d, 28d, 92d, 62d, 39d and 260d (seventh one having period of
17d) respectively. They further showed that this system is dynamically
stable. All these planet candidates have relatively low masses ($\sim$
few $M_{\earth}$) with GJ667Cc, e and f lying inside the habitable
zone, which if confirmed, would make GJ667C one of the first systems
with multiple low mass planets in its habitable zone. They also
considered a model with correlated noise (modified ARMA model
described in \citealt{2013A&A...551A..79T} and Sec.~\ref{sec:RV_like}
of this paper) but found that white noise model is favoured by the
data.

It has already been shown that noise in photometric observations of exoplanetary transits is often correlated \citep{2006MNRAS.373..231P}. Information content of correlated data is lower than if the data were uncorrelated, therefore ignoring correlated noise components can result in spurious detection. In this paper, we present a Bayesian method for the analysis of RV data-sets with correlated noise. We also allow for the possibility of the reported uncertainty values on RV measurement to be over or under-estimated and deal with any inconsistencies between different data-sets in a statistically robust manner. We apply this method to the RV data-set of GJ667C.

The outline of this paper is as follows. We give a brief introduction
to Bayesian inference in Sec.~\ref{sec:bayesian} and describe our
object detection method for calculating the number of planets favoured
by the data in Sec.~\ref{sec:object_detection}. Our method for
modelling RV data is described in Sec.~\ref{sec:RV}. In
Sec.~\ref{sec:bayes_RV} we describe our Bayesian analysis methodology
including the likelihood function and choice of prior
distributions. We apply our method to RV data sets of GJ667C in
Sec.~\ref{sec:results} and present our conclusions in
Sec.~\ref{sec:conclusions}.

%%%%%%%%%%%%%%%%%%%%%%%%%%%%%%%%%%%%%%%%%%%%%%%%%%%%%%%%%
\section{Bayesian inference}\label{sec:bayesian}
%%%%%%%%%%%%%%%%%%%%%%%%%%%%%%%%%%%%%%%%%%%%%%%%%%%%%%%%%

Bayesian inference provides a consistent approach to the estimation of
a set of parameters $\boldsymbol \Theta$ in a model (or hypothesis)
$H$ for the data $\boldsymbol D$. Bayes' theorem states that
\begin{equation} 
\Pr(\boldsymbol \Theta|\boldsymbol D, H) = \frac{\Pr(\boldsymbol D|\,\boldsymbol \Theta,H)\Pr(\boldsymbol \Theta|H)}{\Pr(\boldsymbol D|H)},
\label{eq:Bayes}
\end{equation}
where $\Pr(\boldsymbol \Theta|\boldsymbol D, H) \equiv P(\boldsymbol
\Theta|\boldsymbol D)$ is the posterior probability distribution of
the parameters, $\Pr(\boldsymbol D|\boldsymbol \Theta, H) \equiv
\mathcal{L}(\boldsymbol \Theta)$ is the likelihood, $\Pr(\boldsymbol
\Theta|H) \equiv \pi(\boldsymbol \Theta)$ is the prior, and
$\Pr(\boldsymbol D|H) \equiv \mathcal{Z}$ is the Bayesian evidence
given by:
\begin{equation}
\mathcal{Z} = \int{\mathcal{L}(\boldsymbol \Theta)\pi(\boldsymbol
  \Theta)}d^N\boldsymbol \Theta,
\label{eq:Z}
\end{equation} 
where $N$ is the dimensionality of the parameter space. Bayesian
evidence being independent of the parameters, can be ignored in
parameter estimation problems and inferences can be obtained by taking
samples from the (unnormalized) posterior distribution using standard
MCMC methods.

Model selection between two competing models $H_{0}$ and $H_{1}$ can
be done by comparing their respective posterior probabilities given
the observed data-set $\boldsymbol D$, as follows
\begin{equation}
R = \frac{\Pr(H_{1}|\boldsymbol D)}{\Pr(H_{0}|\boldsymbol D)}
  = \frac{\Pr(\boldsymbol D|H_{1})\Pr(H_{1})}{\Pr(\boldsymbol D| H_{0})\Pr(H_{0})}
  = \frac{\mathcal{Z}_1}{\mathcal{Z}_0} \frac{\Pr(H_{1})}{\Pr(H_{0})},
\label{eq:R}
\end{equation}
where $\Pr(H_{1})/\Pr(H_{0})$ is the prior probability ratio for the
two models, which can often be set to unity in situations where there
is not a prior reason for preferring one model over the other, but
occasionally requires further consideration. It can be seen from
(\ref{eq:R}) that the Bayesian evidence plays a central role in
Bayesian model selection.

As the average of the likelihood over the prior, the evidence is
larger for a model if more of its parameter space is likely and
smaller for a model with large areas in its parameter space having low
likelihood values, even if the likelihood function is very highly
peaked. Thus, the evidence automatically implements Occam's razor.

Evaluation of the multidimensional integral in (\ref{eq:Z}) is a
challenging numerical task. Standard techniques like thermodynamic
integration are extremely computationally expensive which makes
evidence evaluation at least an order of magnitude more costly than
parameter estimation. Various alternative information criteria for
astrophysical model selection are discussed by \citet{liddle07}, but
the evidence remains the preferred method.

The nested sampling approach, introduced by \citet{skilling04}, is a
Monte Carlo method targeted at the efficient calculation of the
evidence, but also produces posterior inferences as a
by-product. \citet{feroz08, multinest, 2013arXiv1306.2144F} built on
this nested sampling framework and have introduced the
{\sc MultiNest} algorithm which is very efficient in sampling from
posteriors that may contain multiple modes and/or large (curving)
degeneracies and also calculates the evidence. This technique has
greatly reduces the computational cost of Bayesian parameter
estimation and model selection and has already been applied to several
inference problems in astro and particle physics (see
e.g. \citealt{2008arXiv0810.0781F, 2009MNRAS.398.2049F,
  2009MNRAS.400.1075B, 2009CQGra..26u5003F, 2011JHEP...03..012B,
  2012arXiv1212.2636S, 2012arXiv1207.3708K}).

%%%%%%%%%%%%%%%%%%%%%%%%%%%%%%%%%%%%%%%%%%%%%%%%%%%%%%%%%
\section{Bayesian Object Detection}\label{sec:object_detection}
%%%%%%%%%%%%%%%%%%%%%%%%%%%%%%%%%%%%%%%%%%%%%%%%%%%%%%%%%

To detect and characterise an unknown number of objects in a data-set,
one would ideally like to infer simultaneously the full set of
parameters $\boldsymbol \Theta = \{N_{\rm obj},\boldsymbol{\Theta}_1, 
\boldsymbol{\Theta}_2, \cdots,
\boldsymbol{\Theta}_{N_{\rm obj}}, \boldsymbol \Theta_{\rm n}\}$, where $N_{\rm obj}$ is the
(unknown) number of objects, $\boldsymbol{\Theta}_{i}$ are the parameters values
associated with the $i$th object, and $\boldsymbol{\Theta}_{\rm n}$ is the set of
(nuisance) parameters common to all the objects. This, however,
requires any sampling based approach to move between spaces of
different dimensionality as the length of the parameter vector depends
on the unknown value of $N_{\rm obj}$. Such techniques are discussed in
\cite{Hobson03} and \cite{2013AJ....146....7B}. Nevertheless, due to
this additional complexity of variable dimensionality, these
techniques are generally extremely computationally intensive.

An alternative approach for achieving virtually the same result is the
`multiple source model'. By considering a {\em series} of models
$H_{N_{\rm obj}}$, each with a {\em fixed} number of objects,
i.e. with $N_{\rm obj}=0,1,2,\ldots$.  One then infers $N_{\rm obs}$
by identifying the model with the largest marginal posterior
probability $\Pr(H_{N_{\rm obj}}|\bmath{D})$. Assuming that there are
$n_{\rm p}$ parameters per object and $n_{\rm n}$ (nuisance)
parameters common to all the objects, for $N_{\rm obj}$ objects, there
would be $N_{\rm obj}n_{\rm p}+n_{\rm n}$ parameters to be inferred,
Along with this increase in dimensionality, the complexity of the
problem also increases with $N_{\rm obj}$ due to the exponential
increase in the number of modes as a result of counting degeneracy
(there are $n!$ more modes for $N_{\rm obj} = n$ than for $N_{\rm obj}
= 1$).

If the contributions to the data from each object are reasonably well
separated and the correlations between parameters across objects is
minimal, one can use the alternative approach of `single source model'
by setting $N_{\rm obj} = 1$ and therefore the model for the data
consists of only a single object. This does not, however, restrict us
to detecting only one object in the data. By modelling the data in
such a way, we would expect the posterior distribution to possess
numerous peaks, each corresponding to the location of one of the
objects. Consequently the high dimensionality of the problem is traded
with high multi-modality in this approach, which, depending on the
statistical method employed for exploring the parameter space, could
potentially simplify the problem enormously. For an application of
this approach in detecting galaxy cluster from weak lensing data-sets
see \cite{2008arXiv0810.0781F}.

Calculating Bayesian evidence accurately for large number of objects
is extremely difficult, due to the increase in dimensionality and severe
complexity of the posterior, but parameter estimation can still
be done accurately. In order to circumvent this problem,
\cite{2011MNRAS.415.3462F} proposed a new general approach to Bayesian
object detection called the `residual data model' that is applicable even
for systems with a large number of planets. This method is based on
the analysis of residual data after detection of $N_{\rm obj}$
objects. We summarize this method below.

Let $H_{N_{\rm obj}}$ denote a model with $N_{\rm obj}$ objects. The
observed (fixed) data is denoted by $\bmath{D} = \{d_1, d_2, \cdots,
d_{\rm M}\}$, with the associated uncertainties being $\{\sigma_1,
\sigma_2, \cdots, \sigma_{\rm M}\}$. In the general case that $N_{\rm
  obj} = n$, the random variable $\bmath{D}_n$ is defined as a
realisation of the data that would be collected if the model $H_n$
were correct, and the random variable $\bmath{R}_n\equiv
\bmath{D}-\bmath{D}_n$, as the corresponding data residuals in this
case. If one analyses the observed data $\bmath{D}$ to obtain
samples from the posterior distribution of the model parameters
$\Theta$, it is straightforward to obtain samples from the
posterior distribution of the data residuals $\bmath{R}_n$. This is
given by
\begin{equation}
\Pr(\bmath{R}_n|\bmath{D},H_n) = \int \Pr(\bmath{R}_n|
\boldsymbol{\Theta},H_n)\Pr(\boldsymbol{\Theta}|\bmath{D},H_n)
\,d\boldsymbol{\Theta},
\label{eqn:residdef}
\end{equation}
where 
\begin{equation}
\Pr(\bmath{R}_n|\boldsymbol{\Theta},H_n) = 
\prod_{i=1}^M \frac{1}{\sqrt{2\pi\sigma_i^2}}\exp\left\{-\frac{[D_i-R_i-D_{{\rm p},i}(\boldsymbol{\Theta})]^2}{2\sigma_i^2}\right\},
\label{eqn:residdef2}
\end{equation}
and $\bmath{D}_{\rm p}(\boldsymbol\Theta)$ is the (noiseless)
predicted data-set corresponding to the parameter values
$\boldsymbol\Theta$. Assuming that the residuals are independently
Gaussian distributed with mean $\langle\bmath{R}_n\rangle = \{r_1,
r_2, \cdots, r_{\rm M}\}$ and standard deviations $\{\sigma^\prime_1,
\sigma^\prime_2, \cdots, \sigma^\prime_{\rm M}\}$ obtained from the
posterior samples, $\langle\bmath{R}_n\rangle$ can then be analysed
with $N_{\rm obj} = 0$, giving the `residual null evidence' $Z_{\rm
  r,0}$, which is compared with the evidence value $Z_{\rm r,1}$
obtained by analysing $\langle\bmath{R}_n\rangle$ with $N_{\rm obj} =
1$. The comparison is thus being made between the model $H_0$ that the
residual data does not contain an additional object and the model
$H_1$ in which an additional object is present.

With no prior information about the number of objects in a data-set,
the original data-set $\bmath{D}$ is first analysed with $N_{\rm
  obj}=1$. If, in the analysis of the corresponding residuals data,
$H_1$ is favoured over $H_0$, then the original data $\bmath{D}$ are
analysed with $N_{\rm obj} = 2$ and the same process is repeated. In
this way, $N_{\rm obj}$ is increased in the analysis of the original
data $\bmath{D}$, until $H_0$ is favoured over $H_1$ in the analysis
of the corresponding residual data. The resulting value for $N_{\rm
  obj}$ gives the number of objects favoured by the data. This
approach thus requires the detection and estimation of orbital
parameters for $N_{\rm obj} = n$ model but the Bayesian evidence
needs to be calculated only for the $N_{\rm obj} = 1$ model (and the $N_{\rm
  obj}=0$ model, which is trivial); this reduces the computational
cost of the problem significantly. We use this method for analysing
the RV data-sets in this paper.

%%%%%%%%%%%%%%%%%%%%%%%%%%%%%%%%%%%%%%%%%%%%%%%%%%%%%%%%%
\section{Modelling Radial Velocities}\label{sec:RV}
%%%%%%%%%%%%%%%%%%%%%%%%%%%%%%%%%%%%%%%%%%%%%%%%%%%%%%%%%

Observing planets at interstellar distances directly is extremely
difficult, since the planets only reflect the light incident on them
from their host star and are consequently many times
fainter. Nonetheless, the gravitational force between the planets and
their host star results in the planets and star revolving around their
common centre of mass. This produces doppler shifts in the spectrum of
the host star according to its RV, the velocity along the
line-of-sight to the observer. Several such measurements, usually over
an extended period of time, can then be used to detect extrasolar
planets.

Following the formalism given in \cite{2009MNRAS.394.1936B}, for
$N_{\rm p}$ planets and ignoring the planet-planet interactions, the
RV at an instant $t_{i}$ observed at $j$th observatory can be
calculated as:
\begin{equation}
v(t_{i},j) = V_{j} - \sum_{p=1}^{N_{\rm p}} K_{p} \left[\sin(f_{i, p} + \varpi_{p}) 
+ e_{p} \sin(\varpi_{p})\right],
\label{eq:RV}
\end{equation}
where
\begin{eqnarray*}
V_{j} & = & \mbox{systematic velocity with reference to $j$th observatory},\\
K_{p} & = & \mbox{velocity semi-amplitude of the $p$th planet},\\
\varpi_{p} & = & \mbox{longitude of periastron of 
the $p$th planet},\\
f_{i, p} & = & \mbox{true anomaly of the $p$th planet},\\
e_{p}& = & \mbox{orbital eccentricity of the $p$th planet},\\ [-1mm]
& & \mbox{start of data taking, at which periastron occurred.}
\end{eqnarray*}
Note that $f_{i, p}$ is itself a function of $e_{p}$, the orbital
period $P_{p}$ of the $p$th planet, and the fraction $\chi_{p}$ of an
orbit of the $p$th planet, prior to the start of data taking, at which
periastron occurred.  While there is a unique mean line-of-sight
velocity of the center of motion, it is important to have a different
velocity reference $V_{j}$ for each observatory/spectrograph pair,
since the velocities are measured differentially relative to a
reference frame specific to each observatory.

Occasionally, there is a long-term linear drift in the RV data owing
to the presence of a distant stellar companion. In such cases, one adds
a corresponding linear drift term to (\ref{eq:RV}) as follows:
\begin{equation}
v(t_{i},j) = V_{j} - \sum_{p=1}^{N_{\rm p}} K_{p} \left[\sin(f_{i, p} + \varpi_{p}) 
+ e_{p} \sin(\varpi_{p})\right] + g(t_{i}-t_0),
\label{eq:RV_2}
\end{equation}
where $g$ is the drift acceleration and $t_0$ is the time of first RV
observation.

\begin{table*}
\begin{center}
\begin{tabular}{|c|c|c|c|c|}
\hline
Parameter & Prior & Mathematical Form & Lower Bound & Upper Bound \\ 
\hline\hline
$P$ (days) & Jeffreys & $\frac{1}{P \ln(P_{\rm max}/P_{\rm min})}$ & $0.2$ & $365,000$ \\
$K$ (m/s) & Mod. Jeffreys & $\frac{(K+K_0)^{-1}}{\ln(1+(K_{\rm max}/K_0)(P_{\rm min}/P_{\rm i})^{1/3}(1/\sqrt{1-e_{\rm i}^2}))}$ & $0$ & $K_{\rm max}(P_{\rm min}/P_{\rm i})^{1/3}(1/\sqrt{1-e_{\rm i}^2})$\\
$V$ (m/s) & Uniform & $\frac{1}{V_{\rm min}-V_{\rm max}}$ & $-K_{\rm max}$ & $K_{\rm max}$ \\
$e$ & Uniform & $1$ & $0$ & $1$ \\
$\varpi$ (rad) & Uniform & $\frac{1}{2\pi}$ & $0$ & $2\pi$ \\
$\chi$ & Uniform & $1$ & $0$ & $1$ \\
$s$ (m/s) & Mod. Jeffreys & $\frac{(s+s_0)^{-1}}{\ln(1+s_{\rm max}/s_0)}$ & $0$ & $K_{\rm max}$ \\
$\alpha$ & Exponential & $e^{-\alpha}$ & 0 & $\infty$ \\
$\tau$ (days) & Uniform & $1$ & 0 & 100 \\ \hline
\end{tabular}
\end{center}
\caption{Prior probability distributions.}
\label{tab:priors}
\end{table*}

The measurement uncertainties on the RV data are assumed to be
uncorrelated and Gaussian-distributed.  In order to allow, however,
for the possibility that the quoted measurement uncertainties are over- or
under-estimated, we introduce a hyper-parameter $\alpha_{j}$, for each
observatory. The uncertainty on $i$th RV measurement from $j$th
observatory, $\sigma_{i,j}$ is modified to become
$\sigma_{i,j}/\alpha_{j}$. As discussed in
\citet*{2002MNRAS.335..377H}, these hyper-parameters effectively assign
a weight to each data-set that is determined directly by its own
statistical properties, and which are then marginalized over. This approach
allows for the consistent statistical analysis of multiple data-sets
even when they would otherwise be mutually inconsistent assuming the
quoted measurement uncertainties. This contrasts sharply with the
common subjective practice of simply excluding certain data-sets
altogether, thereby assigning them a weight of zero.

In order to model the possible presence of an additional correlated
noise component between RVs, which also simultaneously allows us to
model intrinsic stellar variability (`jitter'), we adopt the red noise
model of \cite{2011CeMDA.111..235B, 2013MNRAS.429.2052B}. This
approach is equivalent to assuming the presence of an additional term
$s(t_i,j)$ on the right-hand side of (\ref{eq:RV}) or (\ref{eq:RV_2})
that has a covariance function given by
\begin{equation}
R[s(t_{i},j), s(t_{i^{\prime}},j^{\prime})] = 
s_{j}^{2}\delta_{jj^\prime}\exp(-|t_{i} - t_{i^{\prime}}| / \tau_{j}),
\label{eq:red_covariance}
\end{equation}
where $\delta_{jj'}$ is the Kronecker delta symbol and $\tau_{j}$ is
an unknown parameter characterising the correlation timescale for the
$j$th observatory. For large enough $\tau_{j}$,
(\ref{eq:red_covariance}) becomes:
\begin{equation}
R[s(t_{i},j), s(t_{i^{\prime}},j^{\prime})] = 
s_{j}^{2} \delta_{jj^\prime}\delta_{ii'},
\label{}
\end{equation}
which is the often used `jitter' noise model with no correlated
component. It is worth noting, however, that the correlated noise
component modelled by (\ref{eq:red_covariance}) is generic and need
not arise from intrinsic stellar variability. Indeed, the standard
white noise model should be considered as nested within the red noise
model used in this work.

Therefore, in our model for the RV data, we have five free parameters
$K$, $\varpi$, $e$, $P$ and $\chi$ for each planet, and an additional
linear drift acceleration parameter $g$ when there is linear drift in
the data, common to all the planets.  In addition to these parameters
there are four nuisance parameters $V_{\rm j}$, $s_{j}$, $\alpha_{j}$
and $\tau_{j}$ per observatory. The orbital parameters can be used
along with the stellar mass $m_{\rm s}$ to calculate the length $a$ of
the semi-major axis of the planet's orbit around the centre of mass
and the planetary mass $m$ as follows:
\begin{eqnarray}
a_{\rm s}\sin i & = & \frac{K P \sqrt{1-e^2}}{2 \pi}, \\
\label{eq:as}
m \sin i & \approx &\frac{Km_{\rm s}^{\frac{2}{3}} P^{\frac{1}{3}} \sqrt{1-e^2}}
{(2\pi G)^\frac{1}{3}},
\label{eq:mp}\\
a & \approx & \frac{m_{\rm s} a_{\rm s} \sin i}{m\sin i},
\label{eq:a}
\end{eqnarray}
where $a_{\rm s}$ is the semi-major axis of the stellar orbit about
the centre-of-mass and $i$ is the angle between the direction normal
to the planet's orbital plane and the observer's line of sight. Since
$i$ cannot be measured with RV data, only a lower bound on the
planetary mass $m$ can be estimated.

%%%%%%%%%%%%%%%%%%%%%%%%%%%%%%%%%%%%%%%%%%%%%%%%%%%%%%%%%
\section{Bayesian Analysis of Radial Velocity Measurements}\label{sec:bayes_RV}
%%%%%%%%%%%%%%%%%%%%%%%%%%%%%%%%%%%%%%%%%%%%%%%%%%%%%%%%%

There are several RV search programmes looking for extrasolar
planets. The RV measurements consist of the time $t_{i}$ of the $i$th
observation, the measured RV $v_{i}$ relative to a reference frame and
the corresponding measurement uncertainty $\sigma_{i}$. These RV
measurements can be analysed using Bayes' theorem given in
(\ref{eq:Bayes}) to obtain the posterior probability distributions of
the model parameters discussed in the previous section. We now
describe the form of the likelihood and prior probability
distributions.

%========================================================
\subsection{Likelihood function}\label{sec:RV_like}
%========================================================

As discussed in \cite{2007MNRAS.374.1321G}, the errors on RV
measurements can be treated as Gaussian and therefore the likelihood
function can be written as
\begin{equation}
\mathcal{L}(\Theta) = \frac{1}{|2\pi\mathbfss{C}|^{1/2}} 
\exp\left[-{\textstyle\frac{1}{2}}(\bmath{v}-\bmath{v}^{\prime})^{\rm t}\mathbfss{C}^{-1}(\bmath{v}-\bmath{v}^{\prime})\right],
\end{equation}
where $\bmath{v}$ is the vector with RV measurements $v(t_{\rm i},j)$,
$\bmath{v}^{\prime}$ is the vector with RVs $v(\Theta;t_{\rm i},j)$
calculated using (\ref{eq:RV_2}), and $\mathbfss{C}$ is the covariance
matrix. As discussed above, our model for the RV data includes
hyper-parameters that scale the independent measurement uncertainties
for each observatory and a correlated red noise component in
(\ref{eq:red_covariance}), such that
the total covariance function is given by
\begin{eqnarray}
\mathcal{C}[v(t_{i},j), v(t_{i^{\prime}},j^{\prime})] & & \nonumber \\ 
&& \hspace*{-2cm} = [(\sigma_{i,j}/\alpha_{j})^2 \delta_{ii'}+
s_{j}^{2}\exp(-|t_{i} - t_{i^{\prime}}|/\tau_{j})]\delta_{jj'}.
\end{eqnarray}
This should be contrasted with the common practice when analysing RV
data-sets of adopting a `white' noise model using the quoted
measurement uncertainties directly and ignoring any correlated noise
component, but still including a stellar jitter term, in which case
the covariance function is simply
\begin{equation}
\mathcal{C}_{\rm white}[v(t_{i},j), v(t_{i^{\prime}},j^{\prime})] = 
(\sigma_{i,j}^2 + s_{j}^{2})\delta_{ii'}\delta_{jj'}.
\label{eq:white}
\end{equation}

It should be noted that the red noise model used in this work differs
markedly from the so-called `ARMA' (autoregressive moving-average)
model used in \citet{2013A&A...551A..79T} for modelling the correlated
noise component. The AR part of the ARMA model, with order $p$ models
a time series $X_{i}$ as follows:
\begin{equation}
X_{i} = c + \sum_{i' = 1}^{p} \psi_{i'}X_{i-i'} + \epsilon_{i},
\label{eq:correct_AR}
\end{equation}
where $\psi_{i'}$ are the AR coefficients, $c$ is a constant and
$\epsilon_{i}$ is the white noise term. AR($p$) works well for
regularly spaced time series but since the RV measurements are almost
always irregularly spaced in time, this model in its original form is
not applicable. In order to circumvent this problem,
\citet{2013A&A...551A..79T} modified the AR model given in
(\ref{eq:correct_AR}) as follows:
\begin{equation}
X_{i} = c + \sum_{i' = 1}^{p} \psi_{i,i'}X_{i-i'} + \epsilon_{i},
\label{eq:wrong_AR}
\end{equation}
where
\begin{equation}
\psi_{i,i'} = \psi_{i'}\exp|\gamma(t_{i} - t_{i'})|.
\end{equation}
One potential problem with this approach is that the sampling of time
series at different points in time or at different time resolutions
can have quite a large impact on the way the correlated noise
component is modelled, as the AR($p$) part for calculating $X_{i}$
includes the previous $p$ time series values closest to $X_{i}$,
regardless of their actual temporal separations. The red noise model
that we have adopted correlates every single pair of RV measurements taken
by a given observatory (with the magnitude of correlation dependent on
the temporal separation within the pair) and therefore does not suffer
from this shortcoming.

%========================================================
\subsection{Choice of priors}\label{sec:RV_priors}
%========================================================

For parameter estimation, priors become largely irrelevant once the data are sufficiently constraining, but for
model selection the prior dependence always remains. Therefore, it is important that priors are selected based on
physical considerations. We follow the choice of priors given in \cite{2007MNRAS.374.1321G}, as shown in
Table~\ref{tab:priors}.

The modified Jeffreys prior,
\begin{equation}
\Pr(\theta|H) = \frac{1}{(\theta + \theta_0) \ln(1+\theta_{\rm max}/\theta_0)},
\label{eq:modjeff}
\end{equation}
behaves like a uniform prior for $\theta \ll \theta_0$ and like a Jeffreys prior (uniform in $\log$) for $\theta
\gg \theta_0$. We set $K_0 = s_0 = 1$ m/s and $K_{\rm max} = 2129$ m/s, which corresponds to a maximum
planet-star mass ratio of $0.01$.

The prior distribution imposed on hyper-parameters $\alpha$ is
exponential with expectation value unity. This is because our
expectation is that the uncertainty values on observed RVs are neither
over nor under-estimated, i.e. $E[\alpha] = 1$. With this constraint,
and the fact that each $\alpha$ is a positive quantity, the correct
prior distribution according to the maximum-entropy principle is the
exponential prior (see e.g. \citealt{2002MNRAS.335..377H,Sivia}). When
analysing multiple data-sets jointly, inferred values of
hyper-parameters which are significantly away from unity, may hint at
inconsistency between the data-sets. Nonetheless, inclusion of these
hyper-parameters ensures a statistically consistent analysis of
multiple data-sets even in this case (see
\citealt{2002MNRAS.335..377H} for more details).

%%%%%%%%%%%%%%%%%%%%%%%%%%%%%%%%%%%%%%%%%%%%%%%%%%%%%%%%%
\section{Results}\label{sec:results}
%%%%%%%%%%%%%%%%%%%%%%%%%%%%%%%%%%%%%%%%%%%%%%%%%%%%%%%%%

\begin{figure}
\begin{center}
\includegraphics[width=1.4\columnwidth, angle=-90]{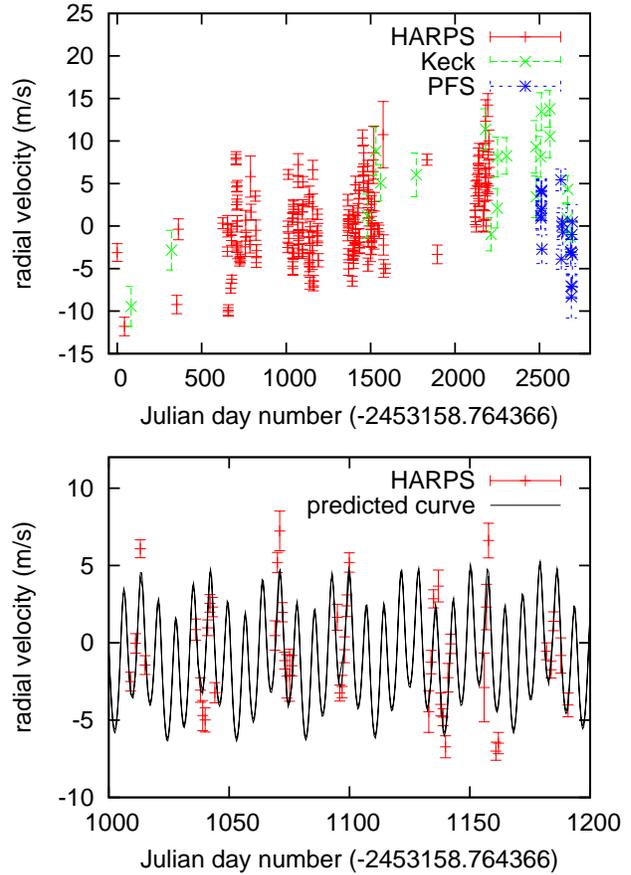}
\caption{Top panel shows the radial velocity measurements for GJ667C
  from $\mathcal{D}_{\rm CCF}$ data-set, with the quoted $1\sigma$
  errorbars. Bottom panel shows a blow-up of the mean fitted radial
  velocity curve to the data for two planets found
  orbiting GJ667C, with the red noise component included in the
  analysis.}
\label{fig:GJ667C_RV}
\end{center}
\end{figure}

\begin{table}
\begin{center}
\begin{tabular}{|c|r|r|r|r|}
\hline
& \multicolumn{2}{c}{$\mathcal{D}_{\rm CCF}$} & \multicolumn{2}{c}{$\mathcal{D}_{\rm TERRA}$}\\
$N_{\rm p}$ &  white noise & red noise &  white noise & red noise\\
\hline\hline
$1$ & $17.05 \pm 0.16$ & $ 4.22 \pm 0.16$ & $16.95 \pm 0.16$ & $ 6.82 \pm 0.16$ \\
$2$ & $ 9.80 \pm 0.16$ & $ 2.24 \pm 0.15$ & $18.94 \pm 0.16$ & $ 5.00 \pm 0.16$ \\
$3$ & $ 2.57 \pm 0.15$ & $ 0.44 \pm 0.14$ & $ 4.22 \pm 0.15$ & $ 0.89 \pm 0.15$ \\
$4$ & $ 0.13 \pm 0.14$ & $ 0.16 \pm 0.14$ & $ 1.37 \pm 0.15$ & $ 0.00 \pm 0.15$ \\ 
$5$ &                  &                  & $-0.49 \pm 0.14$ &                  \\\hline
\end{tabular}
\caption{$\Delta \ln \mathcal{Z_{\rm r}}$ values for the system GJ667C.\label{tab:Z_GJ667C}}
\end{center}
\end{table}

We used the 172 RV measurements of GJ667C obtained by the HARPS
spectrograph with the HARPS-TERRA technique, 20 measurements obtained
with HIRES/Keck and 32 measurements with PFS/Megallan, we call this
data-set $\mathcal{D}_{\rm TERRA}$. We also analysed a separate
data-set called $\mathcal{D}_{\rm CCF}$, containing 170 RV
measurements obtained by HARPS with the CCF technique, along with the
same RV measurements from HIRES/Keck and PFS/Megallan. Both TERRA and
CCF HARPS RV measurements are given in \cite{2013arXiv1306.6074A},
while HIRES/Keck and PFS/Megallan RV measurements are available from
\cite{2012ApJS..200...15A}. Throughout this work, we ignore the
planet-planet interactions and calculate the RVs by assuming Keplerian
orbits for the planets.

The RVs from $\mathcal{D}_{\rm CCF}$ along with their $1-\sigma$
uncertainty values are plotted in the top panel of
Fig.~\ref{fig:GJ667C_RV}. There is an evident long-term linear drift
in RVs of GJ667C induced by its companion stellar pair GJ667AB, with
expected value $\sim 3$ m~s$^{-1}$~yr$^{-1}$ (for a total mass of GJ667AB of 1.27
$M_{\sun}$ and separation between GJ667AB and GJ667C of $\sim 300$ AU)
\citep{2013A&A...553A...8D}. We therefore added an additional drift
component to RVs calculated, as given in (\ref{eq:RV_2}). There is
some hint of correlation between nearby values but due to irregular
temporal sampling, it is difficult to discern any pattern by visual
inspection.

\begin{figure*}
\begin{center}
\includegraphics[width=1.5\columnwidth,height=2\columnwidth]{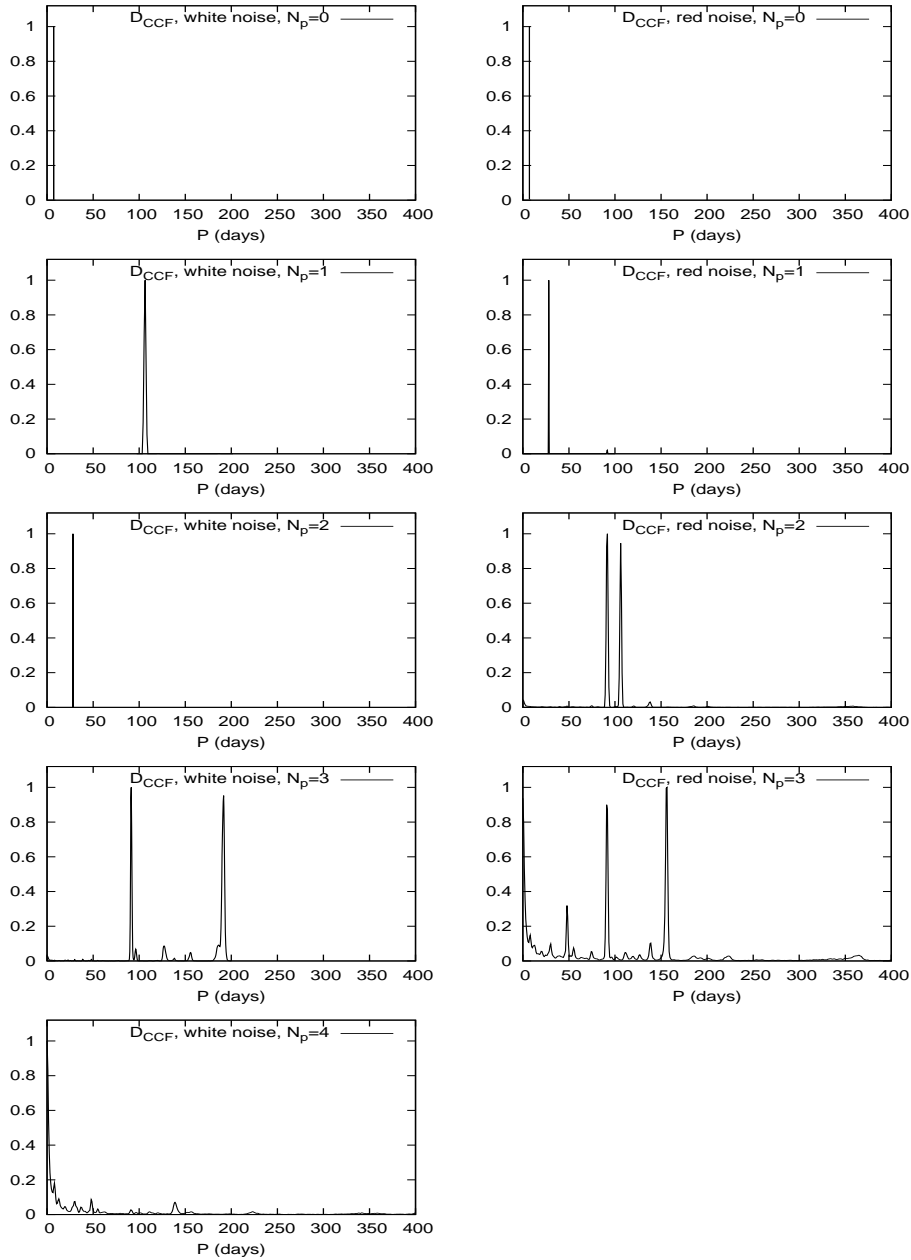}
\caption{1-D marginalised posterior probability distributions for the
  orbital period of planets found in the analysis of residual data
  calculated after the detection of $N_{\rm p}$ planets. Data-set
  $\mathcal{D}_{\rm CCF}$ was used in all cases.}
\label{fig:GJ667Cr-CCF}
\end{center}
\end{figure*}

\begin{figure*}
\begin{center}
\includegraphics[width=1.5\columnwidth,height=2\columnwidth]{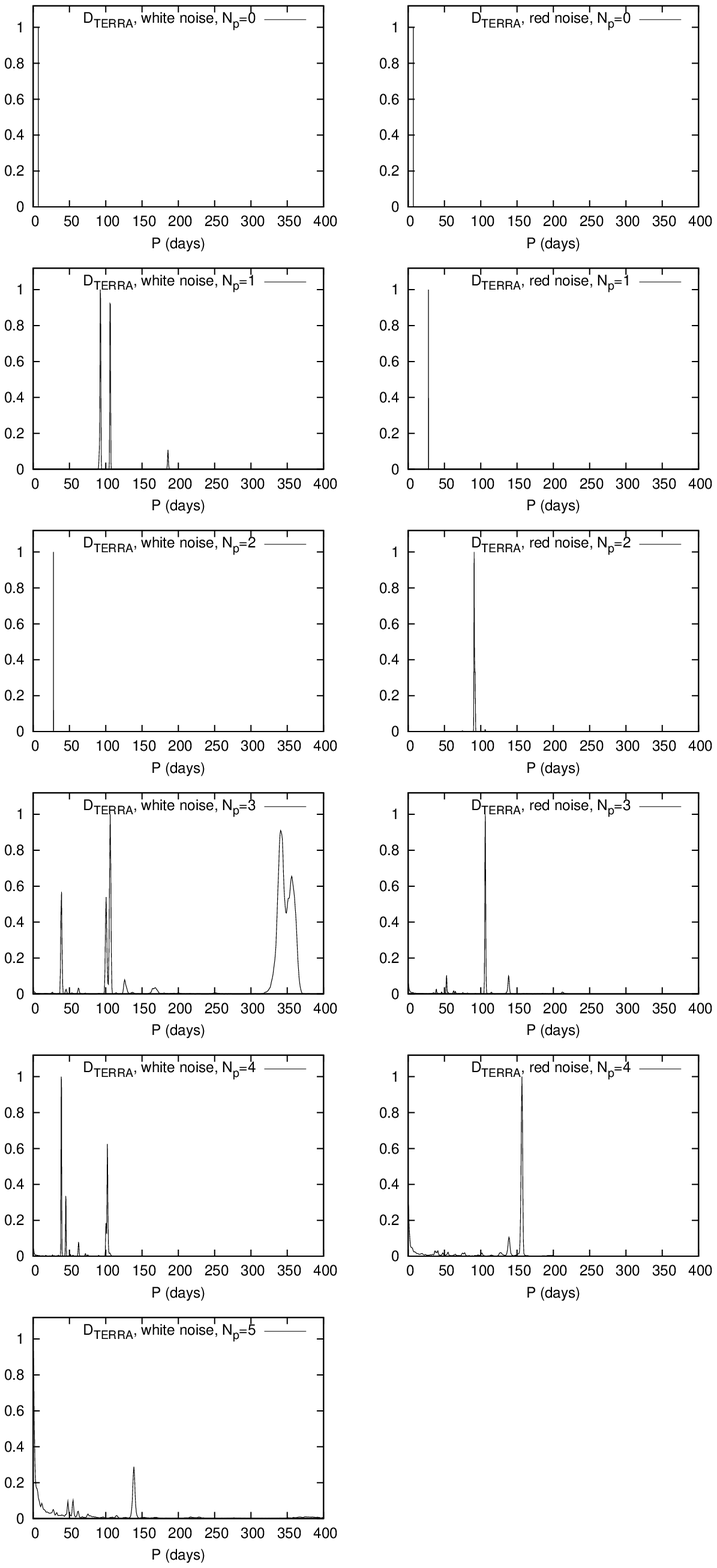}
\caption{1-D marginalised posterior probability distributions for the
  orbital period of planets found in the analysis of residual data
  calculated after the detection of $N_{\rm p}$ planets. Data-set
  $\mathcal{D}_{\rm TERRA}$ was used in all cases.}
\label{fig:GJ667Cr-TERRA}
\end{center}
\end{figure*}

\begin{figure*}
\begin{center}
\includegraphics[width=1.5\columnwidth,height=0.25\columnwidth]{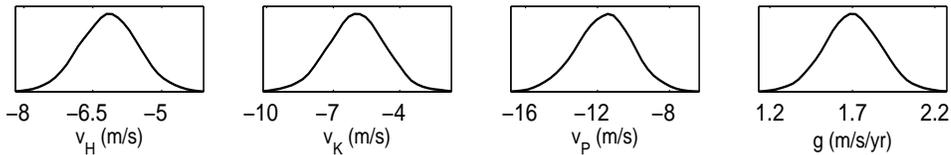}
\caption{1-D marginalised posterior probability distributions for the systematic velocities and drift
acceleration of GJ667C system, obtained by assuming a two planet model, with red noise component included in
analysis of data-set $\mathcal{D}_{\rm CCF}$. Subscripts H, K and P refer to HARPS, Keck and PFS/Magellan 
spectrographs respectively.}
\label{fig:GJ667C_1D_a}
\end{center}
\end{figure*}

\begin{figure*}
\begin{center}
\includegraphics[width=1.5\columnwidth,height=1\columnwidth]{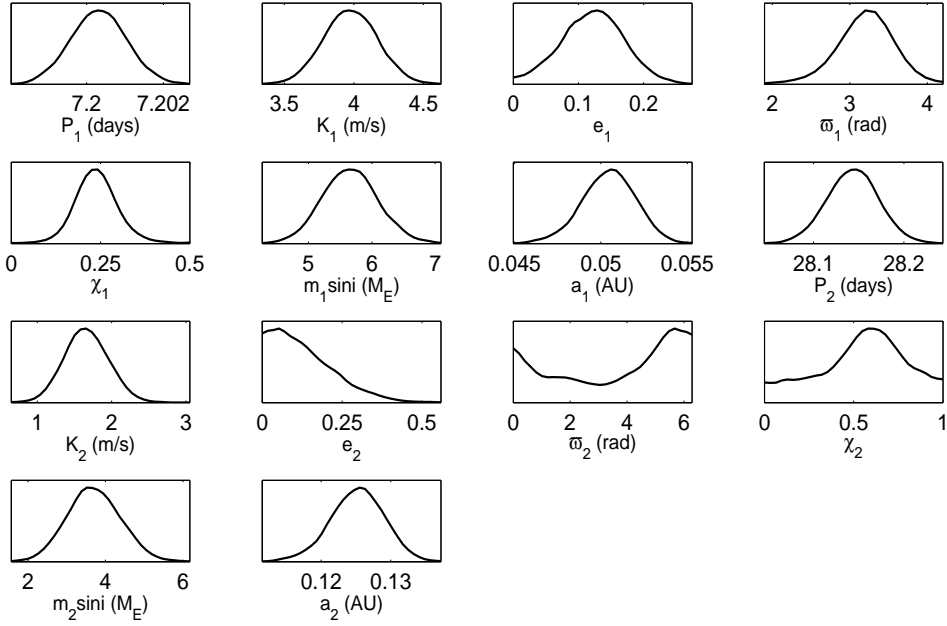}
\caption{1-D marginalised posterior probability distributions for the orbital parameters of the two planets,
found orbiting GJ667C, in analysis of data-set $\mathcal{D}_{\rm CCF}$, with the red noise component included.}
\label{fig:GJ667C_1D_b}
\end{center}
\end{figure*}

\begin{figure*}
\begin{center}
\includegraphics[width=1.5\columnwidth,height=0.75\columnwidth]{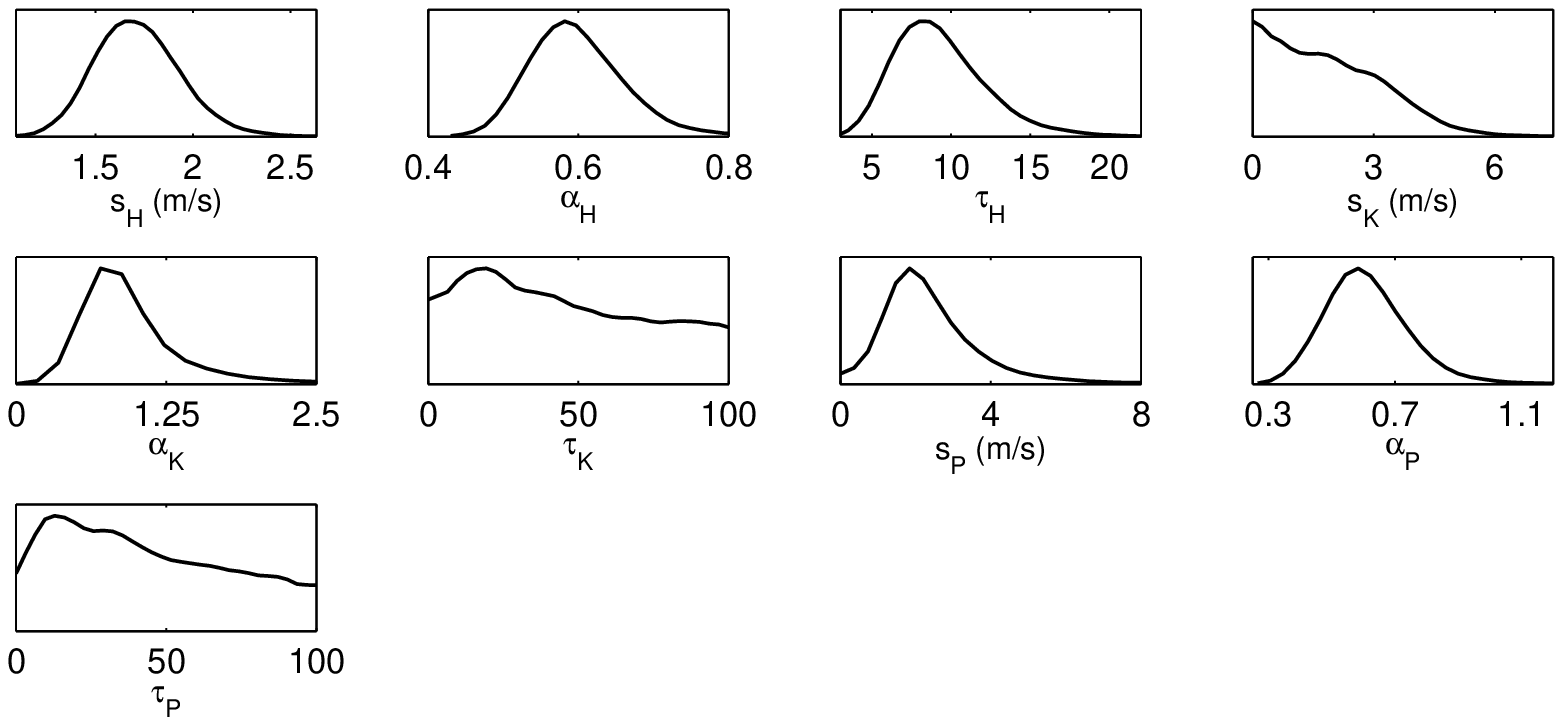}
\caption{1-D marginalised posterior probability distributions for the parameters specific to the noise model,
with red noise component included, obtained by assuming a two planet model in analysis of data-set
$\mathcal{D}_{\rm CCF}$. Subscripts H, K and P refer to HARPS, Keck and PFS/Magellan 
spectrographs respectively.}
\label{fig:GJ667C_1D_c}
\end{center}
\end{figure*}

%\begin{figure*}
%\begin{center}
%\includegraphics[width=1.5\columnwidth,height=1.5\columnwidth]{FIGS/GI667C_corr1_scalenoise12p2.ps}
%\caption{1-D marginalised posterior probability distributions for the
%  parameters of the two planets, with red noise component included in
%  analysis of data-set $\mathcal{D}_{\rm CCF}$, found orbiting
%  GJ667C.}
%\label{fig:GJ667C_1D}
%\end{center}
%\end{figure*}

We first address the question of whether there is evidence for the
presence of correlated noise in the RV data-set of GJ667C. By
comparing the evidence values for models with white and red noise, one
could attempt to answer the question whether this system favours
correlated red noise model over uncorrelated white noise. For the
$\mathcal{D}_{\rm CCF}$ ($\mathcal{D}_{\rm TERRA}$) data-set, $\Delta
\ln \mathcal{Z}$ in favour of red noise for $N_{\rm p} = 0$ and
$N_{\rm p} = 1$ is found to be $18.50 \pm 0.37$ ($19.77 \pm 0.37$) and
$35.83 \pm 0.34$ ($45.46 \pm 0.35$) respectively, clearly showing very
strong evidence in favour of the correlated noise model. Another way
to distinguish between these two noise models would be to determine
whether very large values of correlation timescale $\tau$ are ruled
out when the red noise component is included in the analysis. Looking
ahead, the 1-D marginalised posterior probability distributions for
correlation timescales $\tau_{\rm H}$, $\tau_{\rm K}$ and $\tau_{\rm
  P}$ of HARPS, Keck and PFS, for $N_{\rm p}=2$ planets in the
analysis of data-set $\mathcal{D}_{\rm CCF}$ are shown in
Fig.~\ref{fig:GJ667C_1D_c}. It is clear from these plots that there is a
reasonably tight constraint on $\tau_{\rm H}$ around $\sim$ 9 days
while the posteriors for $\tau_{\rm K}$ and $\tau_{\rm P}$ are largely
unconstrained. Posterior distributions of $\tau$ from the analysis of
data-set $\mathcal{D}_{\rm TERRA}$ are similar. We can therefore be
confident that the HARPS data strongly favours correlated red noise
model over the uncorrelated white noise model. The Keck and PFS
data-sets are not sufficiently discriminative, largely due to not
having enough data points, to rule out either the white or the red
noise model.

The origin of this correlated red noise is not entirely clear. It has 
already been shown that noise in photometric observations of exoplanetary 
transits is often correlated \citep{2006MNRAS.373..231P}. Furthermore, 
\cite{2008MNRAS.386..516O} showed that RV noise is not necessarily white 
due to stellar oscillations. Studies of couple of other M dwarves GJ876 
and GJ581 have also found strong evidence for the presence of red noise 
\citep{2011CeMDA.111..235B, 2013MNRAS.429.2052B}. The correlation 
timescale of order $9$ days found in this study, is too long to be 
explained by stellar oscillations alone and therefore could be due to a
combination of several stellar effects.

In order to determine the number of planets supported by the RV
data-sets of GJ667C, we follow the object detection methodology
outlined in Sec.~\ref{sec:object_detection} and analyse the RV data,
for both the correlated red noise and uncorrelated white noise models,
starting with $N_{\rm p} = 0$ and increasing it until the residual
evidence ratio $\Delta \ln \mathcal{Z_{\rm r}} \simeq 0$. These evidence
ratios, obtained from the residuals after analysing the original data
with a model containing $N_{\rm p}$ planets, are presented in
Table~\ref{tab:Z_GJ667C}.  For each value of $N_{\rm p}$, we also plot
in Figs~\ref{fig:GJ667Cr-CCF} and \ref{fig:GJ667Cr-TERRA} the
corresponding marginalised posterior probability distributions for the
orbital period $P$ obtained from the analysis of the residuals data,
for data-sets $\mathcal{D}_{\rm CCF}$ and $\mathcal{D}_{\rm TERRA}$
respectively. The combination of these residual posterior plots with
the residual evidence values can be viewed as the Bayesian analogue of
the Lomb--Scargle periodogram, with the residual evidences quantifying
the level of confidence in the presence of any additional planets. We
reiterate, however, that in our main object detection
analysis, if $\Delta \ln \mathcal{Z_{\rm r}} \ga 0$ for $N_{\rm p} =
n$, we analyse the {\em original} (rather than residual) data with the
$N_{\rm p} = n+1$ planet model.

For the red noise model, one sees from Table~\ref{tab:Z_GJ667C} that
both $\mathcal{D}_{\rm CCF}$ and $\mathcal{D}_{\rm TERRA}$ show strong
evidence for the presence of no more than three planets. For both
data-sets, the $N_{\rm p}=2$ model yields the planets GJ667Cb and c,
with periods $7.19$d and $28.13$d respectively. For the $N_{\rm p}=3$
model, however, one finds that the third planet has a period of 106d
for $\mathcal{D}_{\rm CCF}$ and 91d for data-set $\mathcal{D}_{\rm
  TERRA}$. Indeed, this is consistent with the posterior distributions
of orbital period from the analysis of residuals data after the
detection of three planets; as shown in Figs~\ref{fig:GJ667Cr-CCF} and
\ref{fig:GJ667Cr-TERRA} these distributions peak at 91d and 106d,
respectively, for data-sets $\mathcal{D}_{\rm CCF}$ and
$\mathcal{D}_{\rm TERRA}$. For the $N_{\rm p}=4$ model, one finds that
all four signals (with periods 7.19d, 28.13d, 91d and 106d) are
detected in both (original) data-sets $\mathcal{D}_{\rm CCF}$ and
$\mathcal{D}_{\rm TERRA}$.

The presence of the 106d signal has already been debated quite extensively (see e.g.
\citealt{2013A&A...553A...8D}), with several studies attributing it to stellar rotation, since it is very close
to the rotation period of the star of 105d. Moreover, the full width at  half-maximum (FWHM) of the CCF, and the
Ca-II H+K S-index in the Mount  Wilson system (S-index), which are used as indicators of stellar  activity, both
show a peak at 105d \citep{2012ApJ...751L..16A}. We also  cannot be sure about the presence of 91d signal, as it
was detected as  the fourth planet in the analysis of $\mathcal{D}_{\rm CCF}$ and the residual evidence for
$N_{\rm p} = 3$ in this case was found to be $\sim$ 0. Furthermore,  both FWHM of the CCF, and the S-index show a
peak at 91d, although the 105d peak in these indicators is much more prominent than the 91d peak
\citep{2012ApJ...751L..16A}. We are therefore confident in our conclusion that the current RV data-set provides
strong evidence only for 2 planets in this system. However, the presence of a third signal with period 91d can
not be ruled out, but the confirmation of its planetary origins will only be possible with more RV observations.
Adopting the two-planet model with the red noise component included, the estimated parameter values obtained from
the analysis of $\mathcal{D}_{\rm CCF}$ are listed in Table~\ref{tab:Z_GJ667C_theta} while the 1-D marginalised
posterior probability distributions are shown in Figs.~\ref{fig:GJ667C_1D_a}-\ref{fig:GJ667C_1D_c}. The mean RV
curve for the two-planet model is overlaid on the RV measurements in Fig.~\ref{fig:GJ667C_RV}. The posterior
distributions obtained from the analysis of $\mathcal{D}_{\rm TERRA}$ are very similar and therefore we do not
re-produce them here.

Assuming the white noise model, one can see from
Table~\ref{tab:Z_GJ667C} that there is evidence for the presence of at
least 4 and perhaps 5 signals depending on whether the
$\mathcal{D}_{\rm CCF}$ or $\mathcal{D}_{\rm TERRA}$ data-set is
used. Apart from the four signals with orbital periods 7.19d, 28.13d,
91d and 106d, there are additional signals with periods 39d, 60d, 180d
and 350d, as can be seen in Figs. ~\ref{fig:GJ667Cr-CCF} and
~\ref{fig:GJ667Cr-TERRA}. Some of these signals have already been
presented as detected planets in several studies (see
e.g. \citealt{2012arXiv1212.4058G, 2013arXiv1306.6074A}). Comparing
the marginalised posterior probability distributions for the orbital
period obtained from the analysis of residual data from white noise
model (Figs.~\ref{fig:GJ667Cr-CCF} and ~\ref{fig:GJ667Cr-TERRA} left
panel) to the red noise model (Figs.~\ref{fig:GJ667Cr-CCF} and
~\ref{fig:GJ667Cr-TERRA} right panel), we can see that there are quite
a few more peaks in the white noise case, showing clear evidence that
erroneously assuming the white noise model leads to spurious
detections of planets. This also gives an explanation for the claims
of detection of up to seven planets in this system.

Finally we note from Fig.~\ref{fig:GJ667C_1D_c} that the hyper-parameter
$\alpha_{\rm H}$, allowing for any under or over-estimation of
measurement uncertainty from the HARPS spectrograph is found to be $0.60
\pm 0.06$ ($0.74 \pm 0.08$) in the analysis of $\mathcal{D}_{\rm CCF}$
($\mathcal{D}_{\rm TERRA}$) data-set, ruling out $\alpha = 1$ (no
under or over-estimation in measurement uncertainties) with high
confidence. Therefore we conclude that the measured uncertainties from
HARPS spectrograph for GJ667C have been under-estimated at the $\sim 50$ 
per cent level.

\begin{table}
\begin{center}
\begin{tabular}{|c|r|r|r|}
\hline
Parameter			& GJ667Cb		& GJ667Cc	     \\
\hline\hline
$P$ (days) 			& $7.200 \pm 0.001$	& $28.143 \pm 0.029$ \\
	 			& $(7.200)$             & $(28.126)$         \\
$K$ (m/s) 			& $3.977 \pm 0.193$	& $1.663  \pm 0.291$ \\
	 		        & $(4.116)$             & $(1.854)$          \\
$e$ 				& $0.122 \pm 0.078$	& $0.133 \pm  0.098$ \\
	 		        & $(0.121)$             & $(0.081)$          \\
$\varpi$ (rad) 			& $3.206 \pm 0.395$	& $3.659 \pm  2.048$ \\
	 		        & $(3.304)$             & $(0.443)$          \\
$\chi$	 			& $0.241 \pm 0.070$	& $0.549 \pm  0.236$ \\
	 		        & $(0.222)$             & $(0.430)$          \\
$m \sin i$ ($M_{\earth}$)	& $5.661 \pm 0.437$	& $3.709 \pm  0.682$ \\
	 		        & $(5.826)$             & $(4.150)$          \\
$a$ (AU)			& $0.050 \pm 0.002$	& $0.125 \pm  0.004$ \\
	 		        & $(0.050)$             & $(0.125)$          \\ \hline
\end{tabular}
\end{center}
\caption{Estimated parameter values for the two planets found orbiting
  GJ667C, with the red noise component included in the analysis of
  data-set $\mathcal{D}_{\rm CCF}$. The estimated values are quoted as
  $\mu \pm \sigma$ where $\mu$ and $\sigma$ are the posterior mean and
  standard deviation respectively. The numbers in parenthesis are the
  maximum-likelihood parameter values.\label{tab:Z_GJ667C_theta}}
\end{table}

%%%%%%%%%%%%%%%%%%%%%%%%%%%%%%%%%%%%%%%%%%%%%%%%%%%%%%%%%
\section{Conclusions}\label{sec:conclusions}
%%%%%%%%%%%%%%%%%%%%%%%%%%%%%%%%%%%%%%%%%%%%%%%%%%%%%%%%%

Detection of extrasolar planets using radial velocity (RV) observations requires the use of statistical model selection techniques. Most of these techniques assume the noise to be uncorrelated. Determining the number of planets from RV data-sets is already a very challenging task due to the problems associated with accurately calculating the probabilities for models with $N_{\rm p} = 0, 1, 2, \cdots$ planets. Allowing for correlated noise adds an additional layer of complexity to this problem. In this work, we have presented a Bayesian method for determining the number of planets supported by RV data-set in the presence of correlated red noise. The red noise model adopted collapses to a white noise model if correlated red noise is not supported by the data. Furthermore, we have introduced hyper-parameters allowing for any over or under-estimation of measurement uncertainties on RV observations. These hyper-parameters also allow us to deal with any inconsistencies between different data-sets in a statistically robust manner. In order to explore the parameter space of these models and perform Bayesian object detection, using the {\sc MultiNest} \citep{feroz08, multinest, 2013arXiv1306.2144F} algorithm whose accuracy has already been demonstrated in many diverse problems in astro and particle physics.

By applying this method to the RV data-set of GJ667C, we find conclusive evidence that the HARPS data favours correlated red noise model over uncorrelated white noise model with the correlation timescale $\sim$ 9 days. Adopting the red noise model, we confirm the presence of planets GJ667Cb and c with periods $7.19$d and $28.13$d respectively. There is some evidence for a third signal with orbital period $91$d, but the planetary origins of this signal are doubtful. We have also shown conclusively that erroneously adopting the white noise model can result in detection of multiple further planets, which also explains the recent claims of the detection of up to seven planets in this system. We also found strong evidence for the under-estimation of measurement uncertainties from the HARPS spectrograph for GJ667C at the $\sim 50$ per cent level which may hint towards some systematics in this data-set.

The level of correlation found in the RV data-set of this system emphasizes the need to check robustly for such correlations before claiming detections of multi-planet systems. This is of vital importance as these multi-planet systems, especially those with planets inside the habitable zone, provide important data for research in many areas of planetary astrophysics.

Finally, we note that although the noise model adopted in this study does a far better job than a white noise model, it is still phenomenological and therefore it does not provide much information about the origin of correlated noise component. One would expect to improve the analysis even further by adopting physically motivated noise models.

%%%%%%%%%%%%%%%%%%%%%%%%%%%%%%%%%%%%%%%%%%%%%%%%%%%%%%%%%
\section*{Acknowledgements}
%%%%%%%%%%%%%%%%%%%%%%%%%%%%%%%%%%%%%%%%%%%%%%%%%%%%%%%%%

This work was performed on COSMOS VIII, an SGI Altix UV1000
supercomputer, funded by SGI/Intel, HEFCE and PPARC, and the authors
thank Andrey Kaliazin for assistance. The work also utilized the
Darwin Supercomputer of the University of Cambridge High Performance
Computing Service (\texttt{http://www.hpc.cam.ac.uk/}), provided by
Dell Inc. using Strategic Research Infrastructure Funding from the
Higher Education Funding Council for England. FF is supported by a
Research Fellowship from the Leverhulme and Newton Trusts. We would
also like to thank the anonymous referee for very useful comments.

\bibliographystyle{mn2e}
\bibliography{references}

\begin{thebibliography}{}

\bibitem[\protect\citeauthoryear{{Anglada-Escude}, {Arriagada}, {Vogt},
  {Rivera}, {Butler}, {Crane}, {Shectman}, {Thompson}, {Minniti},
  {Haghighipour}, {Carter}, {Tinney}, {Wittenmyer}, {Bailey}, {O'Toole},
  {Jones} \& {Jenkins}}{{Anglada-Escude} et~al.}{2012}]{2012ApJ...751L..16A}
{Anglada-Escude} G.,  {Arriagada} P.,  {Vogt} S.~S.,  {Rivera} E.~J.,  {Butler}
  R.~P.,  {Crane} J.~D.,  {Shectman} S.~A.,  {Thompson} I.~B.,  {Minniti} D.,
  {Haghighipour} N.,  {Carter} B.~D.,  {Tinney} C.~G.,  {Wittenmyer} R.~A.,
  {Bailey} J.~A.,  {O'Toole} S.~J.,  {Jones} H.~R.~A.,    {Jenkins} J.~S.,
  2012, \apjl, 751, L16

\bibitem[\protect\citeauthoryear{{Anglada-Escude} \& {Butler}}{{Anglada-Escude}
  \& {Butler}}{2012}]{2012ApJS..200...15A}
{Anglada-Escude} G.,  {Butler} R.~P.,  2012, \apjs, 200, 15

\bibitem[\protect\citeauthoryear{{Anglada-Escude}, {Tuomi}, {Gerlach},
  {Barnes}, {Heller}, {Jenkins}, {Wende}, {Vogt}, {Butler}, {Reiners} \&
  {Jones}}{{Anglada-Escude} et~al.}{2013}]{2013arXiv1306.6074A}
{Anglada-Escude} G.,  {Tuomi} M.,  {Gerlach} E.,  {Barnes} R.,  {Heller} R.,
  {Jenkins} J.~S.,  {Wende} S.,  {Vogt} S.~S.,  {Butler} R.~P.,  {Reiners} A.,
    {Jones} H.~R.~A.,  2013, arXiv e-prints [arXiv:1306.6074]

\bibitem[\protect\citeauthoryear{{Balan} \& {Lahav}}{{Balan} \&
  {Lahav}}{2009}]{2009MNRAS.394.1936B}
{Balan} S.~T.,  {Lahav} O.,  2009, \mnras, 394, 1936

\bibitem[\protect\citeauthoryear{{Baluev}}{{Baluev}}{2011}]{2011CeMDA.111..235B}
{Baluev} R.~V.,  2011, Celestial Mechanics and Dynamical Astronomy, 111, 235

\bibitem[\protect\citeauthoryear{{Baluev}}{{Baluev}}{2013}]{2013MNRAS.429.2052B}
{Baluev} R.~V.,  2013, \mnras, 429, 2052

\bibitem[\protect\citeauthoryear{{Bonfils}, {Delfosse}, {Udry}, {Forveille},
  {Mayor}, {Perrier}, {Bouchy}, {Gillon}, {Lovis}, {Pepe}, {Queloz}, {Santos},
  {S{\'e}gransan} \& {Bertaux}}{{Bonfils} et~al.}{2011}]{2011arXiv1111.5019B}
{Bonfils} X.,  {Delfosse} X.,  {Udry} S.,  {Forveille} T.,  {Mayor} M.,
  {Perrier} C.,  {Bouchy} F.,  {Gillon} M.,  {Lovis} C.,  {Pepe} F.,  {Queloz}
  D.,  {Santos} N.~C.,  {S{\'e}gransan} D.,    {Bertaux} J.-L.,  2011, arXiv
  e-prints [arXiv:1111.5019]

\bibitem[\protect\citeauthoryear{{Brewer}, {Foreman-Mackey} \& {Hogg}}{{Brewer}
  et~al.}{2013}]{2013AJ....146....7B}
{Brewer} B.~J.,  {Foreman-Mackey} D.,    {Hogg} D.~W.,  2013, \aj, 146, 7

\bibitem[\protect\citeauthoryear{{Bridges}, {Cranmer}, {Feroz}, {Hobson}, {Ruiz
  de Austri} \& {Trotta}}{{Bridges} et~al.}{2011}]{2011JHEP...03..012B}
{Bridges} M.,  {Cranmer} K.,  {Feroz} F.,  {Hobson} M.,  {Ruiz de Austri} R.,
   {Trotta} R.,  2011, Journal of High Energy Physics, 3, 12

\bibitem[\protect\citeauthoryear{{Bridges}, {Feroz}, {Hobson} \&
  {Lasenby}}{{Bridges} et~al.}{2009}]{2009MNRAS.400.1075B}
{Bridges} M.,  {Feroz} F.,  {Hobson} M.~P.,    {Lasenby} A.~N.,  2009, \mnras,
  400, 1075

\bibitem[\protect\citeauthoryear{{Delfosse}, {Bonfils}, {Forveille}, {Udry},
  {Mayor}, {Bouchy}, {Gillon}, {Lovis}, {Neves}, {Pepe}, {Perrier}, {Queloz},
  {Santos} \& {S{\'e}gransan}}{{Delfosse} et~al.}{2013}]{2013A&A...553A...8D}
{Delfosse} X.,  {Bonfils} X.,  {Forveille} T.,  {Udry} S.,  {Mayor} M.,
  {Bouchy} F.,  {Gillon} M.,  {Lovis} C.,  {Neves} V.,  {Pepe} F.,  {Perrier}
  C.,  {Queloz} D.,  {Santos} N.~C.,    {S{\'e}gransan} D.,  2013, \aap, 553,
  A8

\bibitem[\protect\citeauthoryear{{Feroz}, {Balan} \& {Hobson}}{{Feroz}
  et~al.}{2011}]{2011MNRAS.415.3462F}
{Feroz} F.,  {Balan} S.~T.,    {Hobson} M.~P.,  2011, Monthly Notices of the
  Royal Astronomical Society, 415, 3462

\bibitem[\protect\citeauthoryear{{Feroz}, {Gair}, {Hobson} \& {Porter}}{{Feroz}
  et~al.}{2009}]{2009CQGra..26u5003F}
{Feroz} F.,  {Gair} J.~R.,  {Hobson} M.~P.,    {Porter} E.~K.,  2009, Classical
  and Quantum Gravity, 26, 215003

\bibitem[\protect\citeauthoryear{{Feroz} \& {Hobson}}{{Feroz} \&
  {Hobson}}{2008}]{feroz08}
{Feroz} F.,  {Hobson} M.~P.,  2008, \mnras, 384, 449

\bibitem[\protect\citeauthoryear{{Feroz}, {Hobson} \& {Bridges}}{{Feroz}
  et~al.}{2009}]{multinest}
{Feroz} F.,  {Hobson} M.~P.,    {Bridges} M.,  2009, \mnras, 398, 1601

\bibitem[\protect\citeauthoryear{{Feroz}, {Hobson}, {Cameron} \&
  {Pettitt}}{{Feroz} et~al.}{2013}]{2013arXiv1306.2144F}
{Feroz} F.,  {Hobson} M.~P.,  {Cameron} E.,    {Pettitt} A.~N.,  2013, arXiv
  e-prints [arXiv:1306.2144]

\bibitem[\protect\citeauthoryear{{Feroz}, {Hobson}, {Zwart}, {Saunders} \&
  {Grainge}}{{Feroz} et~al.}{2009}]{2009MNRAS.398.2049F}
{Feroz} F.,  {Hobson} M.~P.,  {Zwart} J.~T.~L.,  {Saunders} R.~D.~E.,
  {Grainge} K.~J.~B.,  2009, \mnras, 398, 2049

\bibitem[\protect\citeauthoryear{{Feroz}, {Marshall} \& {Hobson}}{{Feroz}
  et~al.}{2008}]{2008arXiv0810.0781F}
{Feroz} F.,  {Marshall} P.~J.,    {Hobson} M.~P.,  2008, arXiv e-prints
  [arXiv:0810.0781]

\bibitem[\protect\citeauthoryear{{Ford}}{{Ford}}{2005}]{2005AJ....129.1706F}
{Ford} E.~B.,  2005, \aj, 129, 1706

\bibitem[\protect\citeauthoryear{{Ford} \& {Gregory}}{{Ford} \&
  {Gregory}}{2007}]{2007ASPC..371..189F}
{Ford} E.~B.,  {Gregory} P.~C.,  2007, in {G.~J.~Babu \& E.~D.~Feigelson} ed.,
  Statistical Challenges in Modern Astronomy IV Vol.~371 of Astronomical
  Society of the Pacific Conference Series, {Bayesian Model Selection and
  Extrasolar Planet Detection}.
pp 189--+

\bibitem[\protect\citeauthoryear{{Gregory}}{{Gregory}}{2005}]{2005ApJ...631.1198G}
{Gregory} P.~C.,  2005, \apj, 631, 1198

\bibitem[\protect\citeauthoryear{{Gregory}}{{Gregory}}{2007}]{2007MNRAS.374.1321G}
{Gregory} P.~C.,  2007, \mnras, 374, 1321

\bibitem[\protect\citeauthoryear{{Gregory}}{{Gregory}}{2012}]{2012arXiv1212.4058G}
{Gregory} P.~C.,  2012, arXiv e-prints [arXiv:1212.4058]

\bibitem[\protect\citeauthoryear{{Hobson}, {Bridle} \& {Lahav}}{{Hobson}
  et~al.}{2002}]{2002MNRAS.335..377H}
{Hobson} M.~P.,  {Bridle} S.~L.,    {Lahav} O.,  2002, \mnras, 335, 377

\bibitem[\protect\citeauthoryear{{Hobson} \& {McLachlan}}{{Hobson} \&
  {McLachlan}}{2003}]{Hobson03}
{Hobson} M.~P.,  {McLachlan} C.,  2003, \mnras, 338, 765

\bibitem[\protect\citeauthoryear{{Karpenka}, {March}, {Feroz} \&
  {Hobson}}{{Karpenka} et~al.}{2013}]{2012arXiv1207.3708K}
{Karpenka} N.~V.,  {March} M.~C.,  {Feroz} F.,    {Hobson} M.~P.,  2013, \mnras

\bibitem[\protect\citeauthoryear{{Liddle}}{{Liddle}}{2007}]{liddle07}
{Liddle} A.~R.,  2007, \mnras, 377, L74

\bibitem[\protect\citeauthoryear{{Lomb}}{{Lomb}}{1976}]{1976Ap&SS..39..447L}
{Lomb} N.~R.,  1976, \apss, 39, 447

\bibitem[\protect\citeauthoryear{{Mackay}}{{Mackay}}{2003}]{MacKay}
{Mackay} D.~J.~C.,  2003, {Information Theory, Inference and Learning
  Algorithms}.
Cambridge University Press, Cambridge, UK

\bibitem[\protect\citeauthoryear{{O'Toole}, {Tinney} \& {Jones}}{{O'Toole}
  et~al.}{2008}]{2008MNRAS.386..516O}
{O'Toole} S.~J.,  {Tinney} C.~G.,    {Jones} H.~R.~A.,  2008, \mnras, 386, 516

\bibitem[\protect\citeauthoryear{{Pont}, {Zucker} \& {Queloz}}{{Pont}
  et~al.}{2006}]{2006MNRAS.373..231P}
{Pont} F.,  {Zucker} S.,    {Queloz} D.,  2006, \mnras, 373, 231

\bibitem[\protect\citeauthoryear{{Scargle}}{{Scargle}}{1982}]{1982ApJ...263..835S}
{Scargle} J.~D.,  1982, \apj, 263, 835

\bibitem[\protect\citeauthoryear{Sivia \& Skilling}{Sivia \&
  Skilling}{2006}]{Sivia}
Sivia D.,  Skilling J.,  2006, Data Analysis A Bayesian Tutorial.
Oxford University Press

\bibitem[\protect\citeauthoryear{{Skilling}}{{Skilling}}{2004}]{skilling04}
{Skilling} J.,  2004, in {Fischer} R.,  {Preuss} R.,   {Toussaint} U.~V.,  eds,
  American Institute of Physics Conference Series {Nested Sampling}.
pp 395--405

\bibitem[\protect\citeauthoryear{{S{\"o}derhjelm}}{{S{\"o}derhjelm}}{1999}]{1999A&A...341..121S}
{S{\"o}derhjelm} S.,  1999, \aap, 341, 121

\bibitem[\protect\citeauthoryear{{Strege}, {Bertone}, {Feroz}, {Fornasa}, {Ruiz
  de Austri} \& {Trotta}}{{Strege} et~al.}{2013}]{2012arXiv1212.2636S}
{Strege} C.,  {Bertone} G.,  {Feroz} F.,  {Fornasa} M.,  {Ruiz de Austri} R.,
   {Trotta} R.,  2013, Journal of Cosmology and Astroparticle Physics, 4, 13

\bibitem[\protect\citeauthoryear{{Tuomi}, {Jones}, {Jenkins}, {Tinney},
  {Butler}, {Vogt}, {Barnes}, {Wittenmyer}, {O'Toole}, {Horner}, {Bailey},
  {Carter}, {Wright}, {Salter} \& {Pinfield}}{{Tuomi}
  et~al.}{2013}]{2013A&A...551A..79T}
{Tuomi} M.,  {Jones} H.~R.~A.,  {Jenkins} J.~S.,  {Tinney} C.~G.,  {Butler}
  R.~P.,  {Vogt} S.~S.,  {Barnes} J.~R.,  {Wittenmyer} R.~A.,  {O'Toole} S.,
  {Horner} J.,  {Bailey} J.,  {Carter} B.~D.,  {Wright} D.~J.,  {Salter} G.~S.,
     {Pinfield} D.,  2013, \aap, 551, A79

\end{thebibliography}

\appendix

\label{lastpage}

\end{document}